\documentclass[prl,twocolumn,nofootinbib,
  superscriptaddress,preprintnumbers,amsmath,amssymb]{revtex4}
\usepackage{graphicx}
\usepackage{color}
\usepackage{amsbsy}
\usepackage{dcolumn}
\usepackage{bm}
\usepackage{upgreek}
\usepackage{epstopdf}
\begin{document}
  \title{The Phase Diagram in Electron-Doped 
    La$_{2-x}$Ce$_x$CuO$_{4-\delta}$}
  \author{H. Saadaoui}
  \email[Corresponding E-mail: ]{hassan.saadaoui@psi.ch}
  \affiliation{Paul Scherrer Institute, Laboratory
    for Muon Spin Spectroscopy, 5232 Villigen PSI, Switzerland}
  \author{Z. Salman}
  \affiliation{Paul Scherrer Institute, Laboratory
    for Muon Spin Spectroscopy, 5232 Villigen PSI, Switzerland}
  \author{H. Luetkens}
  \affiliation{Paul Scherrer Institute, Laboratory
    for Muon Spin Spectroscopy, 5232 Villigen PSI, Switzerland}
  \author{T. Prokscha}
  \affiliation{Paul Scherrer Institute, Laboratory
    for Muon Spin Spectroscopy, 5232 Villigen PSI, Switzerland}
  \author{A. Suter}
  \affiliation{Paul Scherrer Institute, Laboratory
    for Muon Spin Spectroscopy, 5232 Villigen PSI, Switzerland}
  \author{W. A. MacFarlane}
  \affiliation{Department of Chemistry, 
    University of British Columbia, Vancouver, BC V6T 1Z1, Canada}
  \author{Y. Jiang}
  \affiliation{Center for Nanophysics and Advanced Materials, University
    of Maryland, College Park, Maryland 20742, USA}
  \author{K. Jin}
  \affiliation{Beijing National Laboratory for Condensed Matter Physics, 
    Institute of Physics, Chinese Academy of Sciences, Beijing 100190, China}
  \author{R. L . Greene}
  \affiliation{Center for Nanophysics and Advanced Materials, University
    of Maryland, College Park, Maryland 20742, USA}
  \author{E. Morenzoni}
  \affiliation{Paul Scherrer Institute, Laboratory
    for Muon Spin Spectroscopy, 5232 Villigen PSI, Switzerland}
  \author{R. F. Kiefl}
  \email[Corresponding E-mail: ]{kiefl@triumf.ca}
  \affiliation{Department of Physics and Astronomy, 
    University of British Columbia, Vancouver, BC V6T 1Z1, Canada }
  \date{\today}
  \newcommand{\LSCO}{La$_{2-x}$Sr$_x$CuO$_{4-\delta}$}
  \newcommand{\PCCO}{Pr$_{2-x}$Ce$_x$CuO$_{4-\delta}$}
  \newcommand{\NCCO}{Nd$_{2-x}$Ce$_x$CuO$_{4-\delta}$}
  \newcommand{\LCCO}{La$_{2-x}$Ce$_x$CuO$_{4-\delta}$}
  \newcommand{\msr}{$\upmu$SR}
  \newcommand{\lem}{LE-$\upmu$SR}
  \newcommand{\etal}{{\it et al.}}
  \newcommand{\ie}{{\it i.e.}}
  \newcommand{\PRL}[3]{Phys. Rev. Lett. {\bf #1}, {#2} ({#3})}
  \newcommand{\APL}[3]{Appl. Phys. Lett. {\bf #1}, {#2} ({#3})}
  \newcommand{\PRB}[3]{Phys. Rev. B {\bf {#1}}, {#2} ({#3})}
  \newcommand{\PB}[3]{Physica B {\bf {#1}}, {#2} ({#3})}
  \newcommand{\PC}[3]{Physica C {\bf {#1}}, {#2} ({#3})}
  \newcommand{\Nt}[3]{Nature {\bf {#1}}, {#2} ({#3})}
  \newcommand{\Sc}[3]{Science {\bf {#1}}, {#2} ({#3})}
  \newcommand{\RMP}[3]{Rev. Mod. Phys. {\bf {#1}}, {#2} ({#3})}
  \newcommand{\equ}[2]{\begin{equation}\label{#1}{#2}\end{equation}}
  \newcommand{\meq}[2]{\begin{eqnarray}\label{#1}{#2}\end{eqnarray}}
  
  \begin{abstract}
   Superconductors are a striking example of a quantum phenomenon in which electrons move coherently over macroscopic distances without scattering. The high-temperature superconducting oxides (cuprates) are the most studied class of superconductors, composed of two-dimensional CuO$_2$ planes separated  by other layers 
   which control the electron concentration in the planes.  
   A key unresolved issue in cuprates is the relationship  between superconductivity and magnetism. In this paper, we report a sharp  phase boundary of static three-dimensional  magnetic order in the electron-doped superconductor \LCCO\  where  small  changes in doping or depth from the surface switch the material from superconducting to magnetic. Using low-energy spin polarized muons, we find static magnetism disappears close to where superconductivity begins and well below the doping where dramatic changes in the transport properties are reported. These results indicate a higher degree of symmetry between the electron and hole-doped cupratets than previously thought.
  \end{abstract}
  \maketitle

\section{Introduction}
  The electron-doped high-$T_{\rm c}$ superconductors are less well-understood
  than their hole-doped counterparts
  \cite{Bennemann08}. Nevertheless experiments on  \LCCO\ (LCCO),
  \PCCO\ (PCCO) and
  \NCCO\ (NCCO) have great significance in the field  of condensed matter
  physics  because they provide a way  to investigate particle-hole
  symmetry  in  the phase diagram where superconductivity emerges from
  doping  a highly correlated antiferromagnetic Mott insulator  or
  charge transfer insulator \cite{ArmitageRMP10}.  In both $n$ and $p$
type materials, there appears to be a quantum critical transition as a
function  of doping which is characterized by sharp maxima in a
variety of properties near
absolute zero \cite{SachdevPT11}. Remarkably, the superconducting
phases of the hole- and electron-doped materials have the same $d$-wave
pairing symmetry \cite{TsueiPRL00}.  However,  there is uncertainty 
about the position of the critical doping
and even the number  of critical points. This results from the difficulty to
monitor how the antiferromagnetic phase of the parent compound
evolves and changes  as a function of doping, especially on the
electron-doped side \cite{JinNt11,MotoyamaNt07,HelmPRL10}. This issue remains of central
importance  since fluctuations associated  with the  quantum
critical point (QCP) may be  the origin of strong superconducting
pairing and unusual properties in the normal state
\cite{SachdevBook11,Tremblay11}. Although there are many similarities in  the
generic phase diagrams of hole and electron-doped cuprates, there are
also considerable differences \cite{weberNtP10}. In particular,  the magnetism  is much
more prominent on  the electron-doped side and appears to overlap
significantly with superconductivity. Also the
superconducting dome is much smaller and narrower in electron-doped 
systems  \cite{ArmitageRMP10}.  
Both systems  show a pseudo-gap opening below a
temperature $T^*$ in the underdoped region which indicates some kind
of fluctuating charge or spin order
{\cite{MotoyamaNt07,OnosePRB04,ArmitagePRL02,ZimmersEPL05,KyungPRL04}}. 
However it is still unclear if
the character of the pseudo-gap phase is the same in both the electron-
and hole-doped cuprates \cite{ArmitageRMP10}.

As mentioned previously, determining the phase diagram in
electron-doped cuprates is complicated  by the difficulty in
determining  how the magnetism evolves and eventually disappears as a
function of doping.  For example  in  NCCO, inelastic magnetic
neutron-scattering results  show that the long-range antiferromagnetic 
order  disappears close to where superconductivity 
first appears \cite{MotoyamaNt07}. However, both
Shubnikov-de Haas Oscillations \cite{HelmPRL09} and ARPES
measurements \cite{Matsui07} find a transition  well inside
the superconducting dome  where the Fermi surface reconstructs. It is
unclear what happens in between  these two dopings.  One
possibility is that neutrons detect the disappearance of  the long
range 3D antiferromagnetic order where superconductivity begins. However,
another less well ordered phase
persists up to a larger  value of $x_{\rm c}$ where the Fermi surface
reconstructs, leading  to sudden changes in transport properties
\cite{DaganPRL04,DaganPRL05}. LCCO has very similar properties to PCCO
and shows the characteristic $T$-linear 
resistivity below a critical doping of 0.17 and Fermi liquid form ($\rho\propto T^2$) 
above \cite{JinNt11}. Also recent angular magneto-resistance  (AMR)
measurements  reveal evidence for the disappearance of static magnetism
above an optimal doping close  to  $x_{\rm c}=0.14$ where  many transport
properties change abruptly \cite{JinPRB09}.    

\begin{figure}
  \includegraphics[width=1.1\columnwidth]{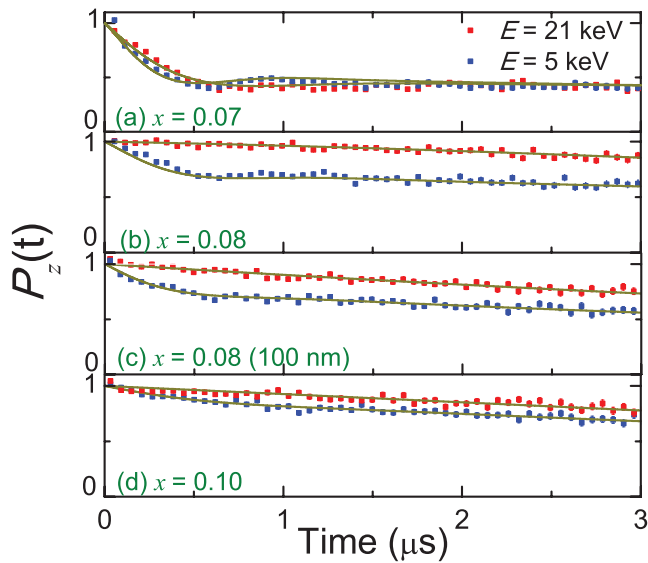}
  \caption{{\bf Typical ZF-\msr\ spectra in \LCCO}. 
    Normalized asymmetry in thin film samples
    with   Ce concentrations of (a) $x=0.07$  of thickness 200 nm, (b) $0.08$ of thickness 200 nm, 
    (c) 0.08 of thickness  100 nm  and, (d) $0.1$ of thickness  200 nm. 
    All spectra are taken in zero applied magnetic field at a temperature of 5 K.  The blue and
    red points correspond to muon implantation  energies of 5 and 21 keV
    respectively. The error bars (too small to see clearly) are statistical uncertainties of the data.
    The solid curves are fits described in Methods.  }
  \label{fig:espectra}
\end{figure}

A major difficulty in determining how the magnetism evolves in 
\LCCO\ is that it can only be made in a thin film form.
Therefore, the traditional experimental techniques for studying
magnetism such as neutron scattering, NMR, and  bulk
$\upmu$SR are not applicable. Recently, the technique of low-energy muon spin
rotation (\lem) has been developed \cite{morenzoni04}. In \lem, the mean  implantation
depth of the muons can be controlled  from a few nm to a few hundred
nm (see Methods). It is now well established that \lem\ is  a
powerful way to investigate both the magnetic and electronic
properties of quantum materials \cite{BorisSc11,SuterPRL11,DunsigerNm10,saadaouiprb13}.

In this paper, we report zero-field (ZF) \lem\ measurements
on thin films of \LCCO\ with  Ce concentrations near the
antiferromagnetic-superconducting boundary.
 The measurements were performed at the $\upmu$E4 beam-line 
of the Swiss Muon Source \cite{prokschaNIM08}, 
at the Paul Scherrer Institute, in Switzerland. 
Samples with Ce concentration of $x=0.07$, 0.08, and 0.10 were
studied and had thicknesses of 200 nm (see Methods). For comparison,  
a fourth film with a Ce  concentration of 
0.08 and  thickness 100 nm  was also studied.
Monte Carlo simulations using TRIM.SP \cite{trim.sp,morenzoniNIMB02} show that 
the average muon implantation depth in \LCCO\ ranges 
from 7  to 110 nm with corresponding straggling of 3 and 23 nm, 
for implantation energies from 1 to 25 keV, respectively.  
Both the mean depth and range straggling are close to being
linear functions of the implantation energy.


\begin{figure}
  \includegraphics[width=\columnwidth]{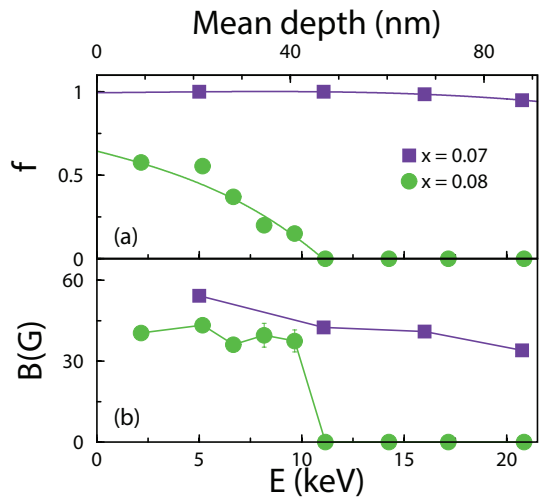}
  \caption{{\bf Energy dependence of the volume fraction and internal field}.
    (a) Magnetic fraction $f$ and (b) average internal 
    magnetic field $B$ as a function of muon implantation energy  at $T=5$ K in LCCO
    films  with  Ce concentration $x=0.07$   and  $0.08$. 
    The $x$-axis  above the top panel
    shows the corresponding mean implantation depth of low-energy muons in LCCO as
    simulated by TRIM.SP.  
    Errors bars give the fit uncertainties, and solid curves are guides to the eye.}
  \label{fig:escan}
\end{figure}
\section{Results}
\subsection{Energy dependence of static magnetism}
Typical ZF-$\upmu$SR spectra at 5 K and energies 5 and 21 keV in all
the  films  are shown in Fig. \ref{fig:espectra}. Note that  at 21 keV, 
corresponding to an average implantation depth of $d\approx 88$ nm and
range straggling $\Delta  \approx 20$ nm, one
can clearly see a fast relaxing signal in the $x=0.07$ sample (red
points in the Fig. \ref{fig:espectra}(a)). Whereas no such fast component is observable
in the two  higher dopings at this energy  (red points in the
Figs. \ref{fig:espectra}(b), Figs. \ref{fig:espectra}(c) and \ref{fig:espectra}(d)).  
This demonstrates that there is bulk static magnetism
in the $x=0.07$ film  which is absent  at the two higher dopings. The
behaviour is markedly different  at the lower energy of 5 keV,
corresponding to $d\approx 22$ nm with $\Delta\approx 20$ nm
(see blue points in  Fig. \ref{fig:espectra}). 
In particular, the fast relaxing signal is  present in both the $0.07$ and both $0.08$ samples, 
but absent in the sample with the $x=0.10$.  The magnetic fraction is slightly larger in the  200 nm thick film compared to the 100 nm film but the transition temperature for the magnetism to disappear  is the same (discussed later).  It
is more subtle but still clear that  the signal is less damped at 5
keV  compared to 21 keV  in the $x=0.07$  sample (red and blue points in
Fig. \ref{fig:espectra}(a)). 
In general, static magnetism is enhanced close to the surface
in all four  samples, with the contrast being largest in the two $x=0.08$ samples. 

The observed relaxation reported here is due to quasi-static magnetic
fields.  This was established with longitudinal field  measurements   
where the applied magnetic field is along the initial spin polarization direction.
In particular, a longitudinal field of 100 G was enough to quench  the observed
relaxation in Figs. \ref{fig:espectra}. This implies the internal magnetic field $B$  must be static  on the time scale of muon Larmor frequency  $\gamma_\upmu B \approx 10^{-7}$s which is typical  of 3D  ordering.

\begin{figure}
  \includegraphics[width=1.1\columnwidth]{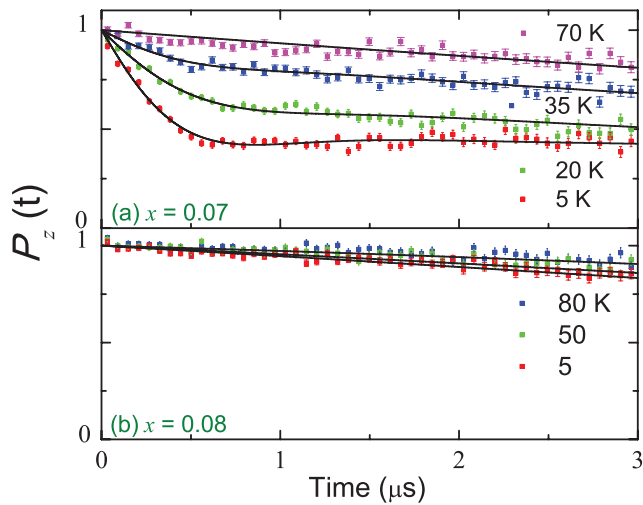}
  \caption{{\bf Typical ZF-\msr\ spectra versus Temperature}. 
    Normalized asymmetry at 21 keV  versus temperature
    in  the bulk of \LCCO\ films with Ce concentrations of  (a) $x=0.07$ and (b) $0.08$. 
    Error bars are statistical uncertainties  of the data, and
    solid curves are fits as described in Methods.}
  \label{fig:tspectra}
\end{figure}

The energy dependence of the magnetic volume fraction $f$ and average internal
field $B$ are shown in Fig. \ref{fig:escan}. The $x=0.07$ sample is fully magnetic
at all energies.   The internal field is about $\sim$60 G at low energy (near the surface) and 
decreases only slightly at the higher energy.
The $x=0.08$ sample appears  magnetic below $\sim$10 keV, where the mean
implantation depth is 40 nm. However, the internal field drops abruptly to zero 
above  10 keV. The results on the  100 nm  thick  film 
  with  $x=0.08$  (not shown here)  are very similar to the 200 nm thick film indicating these affects are
not dependent on the film thickness. In the $x=0.1$ sample there is only a very weak magnetic relaxation 
which is enhanced slightly close to the surface 
(see Fig. \ref{fig:espectra}(d)).

\begin{figure}[t]
  \includegraphics[width=\columnwidth]{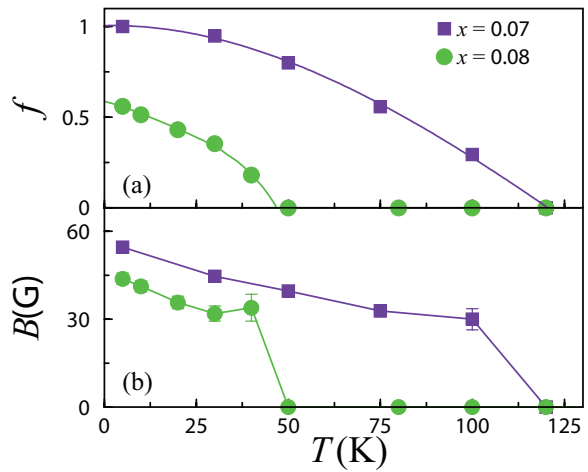}
  \caption{{\bf $T$-dependence of the volume fraction and internal
      field}. (a) Temperature dependence of the magnetic fraction
      $f$, (b) and average internal  magnetic field  $B$; at 5 keV in  LCCO 
    samples with  Ce concentration $x=0.07$  and  $0.08$.  
    Error bars give the fit uncertainties, and solid
    curves are guides to the eye.}
  \label{fig:tscan}
\end{figure}
\subsection{Temperature dependence of static magnetism}
The maximum magnetic ordering temperature in the  region probed by the muons 
is obtained  from the temperature 
dependence  of  ZF-$\upmu$SR spectra,  examples of which  are shown in
Fig. \ref{fig:tspectra}. This shows how the spectrum evolves in
the $x=0.07$ sample going through the magnetic ordering
transition. Below the transition there is a clear evidence for an
over-damped oscillation due to quasi-static  electronic moments which
give rise to a broad distribution of the local internal fields. As  one
approaches the transition, the magnitude of this average field $B$
drops to zero. Above the transition,
in the paramagnetic state,  the electronic moments
are rapidly fluctuating, so the observed weak time dependence of the muon
polarization is due  mostly to  weak quasi-static  nuclear dipolar
fields. In the $x=0.07$ sample at 20 keV (not shown in Fig. 4),
the fitted magnetic fraction $f$  and  the magnitude of  the 
internal field  approach zero near 65 K indicating 
that the entire sample is paramagnetic above this
temperature.  At 5 keV, much closer to the surface,  the
transition temperature is much higher (see  purple points in
Fig. \ref{fig:tscan}). In the $0.08$ sample,  the surface  has  a magnetic
ordering temperature close  to $40(5)$ K  (see  green points in
Fig. \ref{fig:tscan}), but  the bulk is non magnetic
with no sign of static magnetism. This is evident from the fast relaxing signal at low
energies (below 10 keV) as seen earlier in Figs. \ref{fig:espectra}
and \ref{fig:escan}, compared to the slowly relaxing signal at higher
energies. At the highest Ce concentration ($x=0.10$) 
there is no magnetic  ordering at any temperature. It is
important  to note that the muon is sensitive  to any static order both  short  
and long range.  The absence of any static 
magnetic field in a $\upmu$SR experiment (e.g.  in the $x=0.10$ sample)
implies there is no static 3D  magnetism of any type (long range
antiferromagnetism,  spin density wave,  or spin glass).  On the other hand  
it does not exclude quasi  2D  magnetic order in the planes as long as the internal fields are still fluctuating fast compared to  $\gamma_\upmu B \approx 10^{-7}$s. 

\subsection{Phase diagram}
\begin{figure}[t]
   \includegraphics[width=\columnwidth]{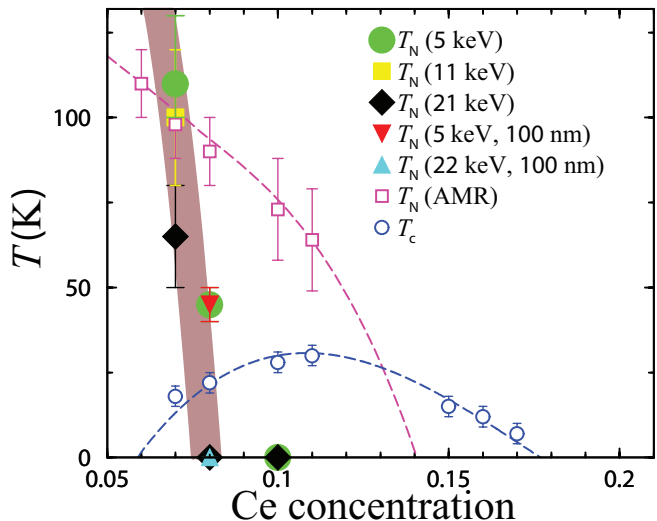}
   \caption{{\bf An updated phase diagram of LCCO}. The magnetic phase
     boundary  measured with \lem\ is the brown band. The  width of  the
     band is due to depth dependence of the magnetism in the film. 
     All data are taken on 200 nm thick films, 
      except one film of 100 nm (triangle symbols).  
     The N$\acute{\rm e}$el temperature ($T_{\rm N}$) from the angular magneto-resistance data is also
     shown for comparison (from Ref. \cite{JinPRB09}). Solid and dashed
     lines are guides to the eye. Error bars of our measurements 
     define the temperature range where magnetism disappears.}
   \label{fig:phase}
 \end{figure}

The results are summarized in the phase diagram in Fig. \ref{fig:phase}.  The
green circles (5 keV), yellow squares  (11 keV), and black diamonds (21 keV) are \lem\ results for 200 nm thick films. 
The red and blue triangles  are for  the 100 nm thick film. 
The brown band defines the magnetic
phase boundary from the \lem\ measurements,  with static magnetism to the
left. The width of the band originates  from the depth dependence of
the magnetism in the film (i.e.  the phase boundary for the bulk  of
the  film is the left side of the brown band).  The dashed blue curve is
the  superconducting transition temperature  obtained from
resistivity, while the cyan dashed curve is the magnetic transition temperature
obtained from angular magneto-resistance (AMR) measurements
\cite{JinPRB09}. 

\section{Discussion}
There is a considerable difference in the
magnetic phase boundary seen with ZF-$\upmu$SR and AMR as reported in Fig. \ref{fig:phase}. In particular
\lem\  shows  a much narrower region of overlap of the static magnetism with
superconductivity. Comparing the results  from these  two techniques  provides an important insight into  the  nature of the magnetism. Angular magneto  resistance (AMR) and  \lem\   are sensitive to magnetic fluctuations  on a  much different time scale. As mentioned above \lem\  measures  the temperature where the internal fields become static on the scale of the period for the Larmor frequency  associated with the internal fields,  which in this case is $10^{-7}$ s.  On the other hand the AMR  detects  magnetism  when the fluctuations become  much slower than the  relaxation time ($10^{-12}$ s). Normally  this would not shift the ordering temperature very much. However LCCO is highly anisotropic so that the coupling between  moments  in the plane  is considerably larger  than  the coupling between planes. Thus  the ordering temperature detected by AMR  is determined  by the  in-plane coupling  between spins $J$. Just below this temperature the moments  are highly correlated  in the CuO$_2$ plane  but  are still  dynamic on the time scale detectable with \lem.   The ordering temperature  in \lem\ defines  the boundary for static 3D ordering, which depends and varies logarithmically with the much smaller magnetic  coupling  between planes $J$'  \cite{YasudaPRL05}. 
This gives new insight into the  relationship between the magnetism in the region of the SC dome  before the 
Fermi surface reconstructs. Apparently  the moments are still   fully developed   above $T_{\rm c}$ but are very dynamic  on the muon time scale. Finally we note that the  \lem\ measurements are performed in zero  applied magnetic field whereas  the AMR measurements are made in a  magnetic field of  14 T  applied in plane. It is possible that the large  magnetic field used in AMR would frustrate the 2D antiferromagnetic order and thereby suppress ordering temperature. This  could explain  why the  AMR 2D phase boundary seems to cross  the \lem\ 3D phase boundary at $x=0.07$.

From the \lem\  results in the bulk of the film,  we conclude that the 
static 3D  static magnetism disappears  close to where superconductivity 
begins and thus  there is little  overlap
region  between static 3D magnetism and superconductivity. 
This is similar to  what is found in \LSCO\ except of the 
evidence of phase separation in bulk, such that a weak spin glass phase
with  a small volume fraction extends into the superconducting
dome in \LSCO\ \cite{Julien03}.  It is possible that a similar spin glass phase
exists in LCCO in bulk but the  glass temperature would have to be lower
than our base temperature of 4 K.
Thus the present  results are consistent with a sharp boundary
and  competition between the superconductivity and magnetism without
 microscopic coexistence. However comparing  these results  with AMR   leads us  to the conclusion that above the SC dome    the moments are fully developed and highly correlated  within the CuO$_2$  plane but  are still fluctuating without any 3D order. Striking similarities between hole and
electron-doped cuprates are also reported by transport studies \cite{JinNt11}.
It is now clear there are  at least two quantum critical points in
\LCCO. The lower one is near $x=0.08$ where the antiferromagnetic disappears, whereas, the
Fermi surface reconstructs  at a much higher doping level above 
$x=0.12$ \cite{HelmPRL10,JinNt11}. It is worth noting that recent studies of certain heavy
fermions show a similar phase diagram with  two critical points  or
possibly  a line of quantum critical points called a quantum critical
phase in between \cite{Coleman2010}.

The difference in the magnetic behaviour of the surface compared
to the bulk of the film is also very interesting. 
It is possible that small structural changes, expected  near a free
surface, are enough to tip the balance towards magnetism. This would be
consistent  with a small difference in (free) energy between competing
phases  near a critical point. As mentioned previously 
 \lem\ measurements were also taken  on a thinner  (100 nm) sample with
 $x=0.08$.  The magnetic properties  of the free surface  were the
 same as in the 200 nm thick  film  with the same value of $x$.
 However  in the thinner film  it was possible, using 21 keV muons,
 to probe the interface between the film and the substrate.   These
 measurements showed this interface was non-magnetic, implying the
 magnetism  is not simply a consequence of the broken translational
 symmetry  and  resulting boundary conditions.   

In conclusion, we have conducted a depth-resolved \lem\ study of
magnetism in \LCCO\ films close to the magnetic-superconducting
transition region.  We find that the near surface region tends to be
more magnetic than the bulk of the films.  The enhanced magnetism in
the near surface region of $x=0.08$ is a property of the free surface
and absent near the substrate interface. The disappearance  of strong static 
magnetic order in the bulk occurs just below  $x=0.08$, which is close
to where superconductivity appears. Above this critical value of $x$  the moments may exist but are fluctuating  rapidly on the muon time scale.  The AMR results there is a high degree of 2D order above the superconducting dome.
Thus above the superconducting transition the system is in    a  dynamic  magnetic state  
with strong antiferromagnetic correlations.   This is similar to the hole-doped 
counterpart \LSCO\ and consistent with a  competition between
order parameters. These results suggest there may be a higher degree
of particle-hole symmetry in this critical region of the phase diagram
than previously thought.

{\bf METHODS}

{\bf \lem\ experiment}: The measurements are performed using low-energy \msr\
technique, where an intense high-energy beam of  muons 
is moderated in a solid Ar film.  A small percentage 
of the incident muons emerge from the argon surface at  low-energy 
($\approx$10 eV). These are subsequently accelerated to 15 keV and transported to
the  sample chamber and  $\upmu$SR spectrometer. The samples are mounted
onto a Ag coated metal plate which is electrically isolated from the
cold  finger of the cryostat and biased to a high voltage ranging
from $-12.5$ to  $12.5$ kV. This allows
the implantation energy of the muons to be adjusted between about $1.5$ and
$26.5$ keV.  The measurements reported here were performed in zero
external magnetic field (ZF) such that  any stray magnetic field at
the sample was less than 0.01 mT. Measurements in a longitudinal
field and transverse field were also performed but are not shown here.
The time evolution of the muon  spin polarization ${P}(t)$, 
which is  measured  through the properties
of the muon  decay, depends on the static and fluctuating components
of the  internal magnetic  field  at the site of the muon \cite{yaouancmuSR10}.  

{\bf Samples}: The $c$-axis-oriented \LCCO\ films were deposited directly 
on insulating (100) SrTiO$_3$ substrates by a pulsed laser deposition
technique utilizing a KrF excimer laser as the exciting light
source. The three films had  Ce concentrations of $x=0.07$, $0.08$ and
$0.10$. Since the oxygen content has an influence on both
the superconducting and normal state properties of the material, 
the annealing process is optimized for each $x$.  
We studied  three  Ce concentration: 0.07
($T_{\rm c}=18\pm2$ K), $0.08$ ($T_{\rm c}=22\pm2$ K), and
0.10 ($T_{\rm c}=27.6\pm0.5$ K).
The sample with $x=0.1$ is metallic (with ${\rm d}\rho/{\rm d}T>$ 0) 
above $T_{\rm c}$ with a narrow transition  width. 
The $x=0.07$ and  $x=0.08$ samples  showed 
an upturn (either in a field or ZF) at low temperature. 
Each \LCCO\ sample consisted  of four identical pieces with a total
area of 2$\times$2 cm$^2$ and 
was attached to the sample holder with a conductive Ag paint.

{\bf Analysis}: The ZF \lem\ 
data is fit  \cite{SuterPP12} to a sum of two functions;
$P_z(t)=P_{\rm LCCO}+P_{\rm Ag}$, corresponding to 
the signal from the fraction of muons stopping in LCCO, and the
remaining fraction of muons ($20-30$ $\%$) stopping in the Ag mounting
plate.
The latter is accounted for with an exponential 
with a small relaxation rate of $0.02-0.04$ $\upmu$s$^{-1}$.  
The signal from the  LCCO may be  decomposed into 
a magnetic fraction $f$ and a non-magnetic or paramagnetic fraction 
$1-f$;
\begin{eqnarray}
P_{\rm LCCO}&=&G_{\rm KT}(t){\rm e}^{-\lambda_{\rm S} t}
 \Big[ f G_{\rm mag} +  (1-f)\Big].
\end{eqnarray}
The magnetic fraction is fit to a phenomenological  function 
consisting of three relaxation 
functions: (1) a weakly relaxing Kubo-Toyabe function  
$G_{\rm KT}(t)=\frac{1}{3}+\frac{2}{3}(1-\sigma_{\rm N}^2
t^2)e^{-\frac{1}{2}\sigma_{\rm N}^2 t^2}$ due to nuclear moments ($\sigma_{\rm N}\approx 0.05-0.1$
$\upmu$s$^{-1}$). 
This term dominates well  above
the ordering temperature  when the electronic moments are rapidly
fluctuating  but has little effect at low temperatures where the
electronic moments dominate. (2) A slowly relaxing exponential,
${\rm e}^{-\lambda_{\rm S} t}$, attributed mostly to slow fluctuations 
in the nuclear dipolar field. (3) A term which
takes into account a broad distribution of large quasi-static internal
magnetic fields from the magnetism; 
$G_{\rm mag}=\frac{1}{3}+\frac{2}{3}\cos(\gamma_B B t)e^{-\lambda_{\rm F} t}$.
For simplicity, we assume the direction of the
internal magnetic field is random in orientation as in a spin glass, giving rise to
the geometric ``$\frac{1}{3}$'' non precessing component appearing in $G_{\rm mag}$. 
The  parameter $B$ represents the average
internal magnetic field from the static ordering, whereas
$\lambda_{\rm F}$ characterizes  the width of this field distribution. This term
dominates at low  temperatures in the ordered state. Since there is an
interplay between $B$ and $\lambda_{\rm F}$, the latter is parametrized as
$B=k\lambda_{\rm F}$, where $k$ is a fitted parameter. This assumes that the
static width of the magnetic field distribution is proportional to the average
field. In the non-magnetic fraction of the sample  
only the first two terms in the relaxation function contribute. 
It should be noted that other fitting functions have been
tested, leading to very similar results.

{\bf Acknowledgments}: 
The authors H.S. and E.M. acknowledge 
the financial support of the MANEP program, and R.F.K. 
the support of NSERC.
The authors J.Y. and R.L.G. acknowledge the support of Maryland Center
for Nanophysics and Advanced Materials and the NSF grant under
DMR-1104256. We would like to thank Johnpierre Paglione 
for  important discussions at the early stages of this work,
and Mohamed Azzouz for commenting and critical reading of this manuscript.

{\bf Contributions}: 
Project planning: R.F.K., H.S., R.L.G., and Z.S.; 
Sample growth: J.P., K.J., and R.L.G.;
\lem\ experiment: H.S., Z.S., R.F.K., E.M., T.P., H.L., A.S., and W.A.M.; 
Data analysis and interpretation: H.S., and R.F.K.;
Draft writing: H.S. and R.F.K., with contributions and/or comments from all authors.

{\bf Competing financial interests}:
The authors declare no competing financial interests.

\end{document}